\documentclass[aps,pre,twocolumn,showpacs,superscriptaddress,letterpaper]{revtex4}

\usepackage{graphicx}
\usepackage{bm}
\usepackage{amsmath}
\usepackage{SIunits}
\usepackage[normalem]{ulem}
\usepackage{color}
\usepackage{xcolor}

\renewcommand{\bm}[1]{\boldsymbol{\mathbf{#1}}}
\newcommand{\ud}{\mathrm{d}}

\newcommand{\vbar}{\bar{v}}

\newcommand{\NoAutoSpaceBeforeFDP}[0]{}
\newcommand{\AutoSpaceBeforeFDP}[0]{}

\begin{document}

\title{Signatures of L\'evy flights with annealed disorder}

\author{Q. Baudouin}
\affiliation{Universit\'e de Nice Sophia Antipolis, CNRS, Institut Non-Lin\'eaire de Nice, UMR 7335, F-06560 Valbonne, France}
\author{R. Pierrat}
\affiliation{ESPCI ParisTech, PSL Research University, CNRS, Institut Langevin, 1 rue Jussieu, F-75005, Paris, France}
\author{A. Eloy}
\affiliation{Universit\'e de Nice Sophia Antipolis, CNRS, Institut Non-Lin\'eaire de Nice, UMR 7335, F-06560 Valbonne, France}
\author{E.J. Nunes-Pereira}
\affiliation{Department of Physics and Center of Physics, University of Minho, 4710-057 Braga, Portugal}
\author{P.-A. Cuniasse}
\affiliation{ESPCI ParisTech, PSL Research University, CNRS, Institut Langevin, 1 rue Jussieu, F-75005, Paris, France}
\author{N. Mercadier}
\affiliation{Universit\'e de Nice Sophia Antipolis, CNRS, Institut Non-Lin\'eaire de Nice, UMR 7335, F-06560 Valbonne, France}
\affiliation{Now at Saint-Gobain Recherche, 39 Quai Lucien Lefranc, 93303 Aubervilliers, France}
\author{R. Kaiser}
\affiliation{Universit\'e de Nice Sophia Antipolis, CNRS, Institut Non-Lin\'eaire de Nice, UMR 7335, F-06560 Valbonne, France}
\email{Robin.Kaiser@inln.cnrs.fr}

\date{\today}

\begin{abstract}
   We present theoretical and experimental results of L\'evy flights of light originating from a random walk of photons
   in a hot atomic vapor. In contrast to systems with quenched disorder, this system does not present any correlations
   between the position and the step length of the random walk. In an analytical model based on microscopic first
   principles including Doppler broadening we find anomalous L\'evy-type superdiffusion corresponding to a single-step
   size distribution $P\left(x\right) \propto x^{-(1+\alpha)}$, with $\alpha\approx1$.  We show that this step size
   distribution leads to a violation of Ohm's law [$T_{\textrm{diff}} \propto L^{-\alpha/2} \neq L^{-1}$], as expected
   for a L\'evy walk of independent steps.  Furthermore the spatial profile of the transmitted light develops power law
   tails [$T_{\textrm{diff}}(r)\propto r^{-3-\alpha}$]. In an experiment using a slab geometry with hot rubidium vapor,
   we measured the total diffuse transmission $T_{\textrm{diff}}$ and the spatial profile of the transmitted light
   $T_{\textrm{diff}}(r)$.  We obtained the microscopic L\'evy parameter $\alpha$ under macroscopic multiple scattering
   conditions paving the way to investigation of L\'evy flights in different atomic physics and astrophysics systems.
\end{abstract}


\maketitle

\section{Introduction}

Transport of waves in complex media has been studied for many decades, even though many fundamental or applicative
questions are still under debate such as imaging through optical thick samples~\cite{SEBBAH-2001} or Anderson
localization~\cite{ANDERSON-1958}.  Light propagation in turbid media can often be described by a diffusion formalism
where photons undergo a random walk process with a step-size distribution $P(x)$ vanishing faster than $1/x^3$ at
infinity. In that case the central limit theorem applies and the mean-square displacement of photons is proportional to
time. This is typically the case for light in cloudy atmospheres~\cite{THOMAS-1999} or in biological
tissues~\cite{TUCHIN-1997}.

However, many physical systems can be subdiffusive or superdiffusive depending on the random walk process
characteristics~\cite{KLAGES-2008}. One particular mechanism for superdiffusion is called L\'evy flights. The
occurrence of L\'evy flights has been investigated in a large variety of systems, ranging from trajectories of
animals~\cite{VISWANATHAN-1996, EDWARDS-2007}, human travel~\cite{BROCKMANN-2006,GONZALEZ-2008},
turbulence~\cite{SHLESINGER-1987}, earthquakes~\cite{CORRAL-2006} to solar flares~\cite{BAIESI-2006}. In optics,
engineered materials where photons trajectories are described by L\'evy flights have been realized recently by stacking
glass spheres with a specific size distribution~(and are thus called L\'evy
glasses)~\cite{WIERSMA-2008-1,BERTOLOTTI-2010}.  In many of these systems, as in most turbid media relevant in
applied physics~\cite{WIERSMA-2013-2}, possible correlations exist between the step-size distribution and the past
history of each random trajectory. These are {\itshape memory effects} giving rise to correlated random walks which
constitute a nuisance on top of the basic L\'evy signature. The origin of the correlations can be traced back either to
the initial position of the step depending on the underlying topography of the system~(e.g.\@ quenched disorder in
L\'evy glasses)~\cite{WIERSMA-2010-3}, or alternatively, to correlations in optical frequency in case of inelastic
scattering~(e.g.\@ partial frequency redistribution effects in atomic vapors)~\cite{NUNES-PEREIRA-2007}. Whatever the
cause, these correlations are at the root of some limitations in the observation of L\'evy flights
characteristics~\cite{BEENAKKER-2012}, and it is thus important to look for a system without any of such correlations.

As suggested early on by Kenty~\cite{KENTY-1932} and reformulated later in the context of L\'evy
flights~\cite{PEREIRA-2004}, superdiffusion of light is also expected in many different atomic systems ranging from
dense atomic vapors, hot plasmas to stars. In this article, we explore theoretically, numerically and experimentally the
propagation of light in a slab containing a hot rubidium gas. This system corresponds to an annealed disorder where no
correlation exists between the step-size distribution and the position~\cite{BEENAKKER-2012}. In an annealed
disordered system, statistical properties do not depend on the position and all positions are strictly identical. In a
quenched system, the position plays a role (not thermal equilibrium) and step correlations have to be taken into
account~\cite{WIERSMA-2010-3,VYNCK-2014-2}. Furthermore, we worked in a regime of so high temperature for the
rubidium vapor~(between $\unit{40}{\celsius}$ to $\unit{170}{\celsius}$) that the assumption of complete frequency
redistribution is valid. This effectively nulify any memory effects associated with inelastic scattering. Additionally,
the range of step-size distribution in atomic vapors is only limited by nearest neighbor distances and by the size of
the sample, allowing for a huge increase in the dynamics of the signals. All of these render the atomic vapor in this
high temperature regime a good experimental system to study supperdifusion, freed from the added complication of
correlated steps sizes. Previous work on L\'evy flights of light have already been conducted in atomic vapors but these
concentrated in the microscopic details: the influence of the microscopic spectral broadening mechanism into the
single-step size distribution~\cite{HOLSTEIN-1947,PEREIRA-2004,KAISER-2009}. All the experimental work done on
macroscopically averaged quantities was only phenomenological. To the best of our knowledge, this is the first report of
an analytical connection between the microscopic L\'evy parameter and the macroscopically measurable observables, for
superdiffusion.

In our theoretical approach, we derive a transport equation from first principles (microscopic approach starting from
Maxwell equations) and with controlled approximations. This equation describes the incoherent propagation of radiation
inside a gas consisting of two-level atoms at high temperature. Solving this integro-differential equation using a Monte
Carlo technique allows us to obtain both the intensity profile $T_{\textrm{diff}}(r)$ at the exit surface of the sample
as well as the total diffuse transmission
\begin{equation}
   T_{\textrm{diff}}=\int_0^{\infty}T_{\textrm{diff}}(r) r\ud r.
\end{equation}
A careful analytical analysis of the integro-differential equation also leads us to an analytical expression for the
step-size distribution $P(x)$, which we use to obtain an analytical prediction for the power law decrease of
$T_{\textrm{diff}}(r)$ which is directly related to the L\'evy exponent $\alpha$.

In an experiment using a slab geometry with hot rubidium atoms, we have measured the total diffuse transmission
$T_{\textrm{diff}}$ and the transmitted intensity profile $T_{\textrm{diff}}(r)$, both in excellent agreement with the
theoretical predictions. The large dynamic range of the experiment data provides a reliable measurement for the L\'evy
exponent $\alpha$, measured in single shot images under multiple scattering conditions and is very close to the result
obtained in a previous study based on the microscopic step-size distribution of light in the system~\cite{KAISER-2009}.

The paper is organised as follows: In Part~\ref{1st_part}, we derive a transport equation for light propagating in hot atomic
vapors and stress several signatures of L\'evy flights encoded in this equation by comparing to the classical case. In
Part~\ref{2nd_part}, we present the experiment in a hot cell of Rubidium gas and show that the experimental results are 
in excellent agreement with the theoretical predictions, for superdiffusion in an annealed disorder system.

\section{Theory}\label{1st_part}

\subsection{Derivation of a transport equation for light in hot atomic gases}

We start the derivation of the transport equation for light in atomic clouds with the polarizability of the atoms
described by
\begin{equation}
   \bm{\alpha}\left(\delta\right)=-\frac{4\pi}{k_0^3}\frac{1}{i+2\delta/\Gamma}
\end{equation}
where $\delta=\omega-\omega_0$ is the detuning, $\omega_0$ the resonance frequency, $\Gamma$ the linewidth and
$k_0=\omega_0/c_0$ the wavevector. This expression is valid for a two-level atom far from saturation and does not take
into account the multilevel structure of the real atoms considered in the experiment. From this building block and from
first principles (Maxwell equations), we derive a transport equation valid for a dilute gas at any temperature. It reads
in the temporal Fourier space (see Ref.~\cite{PIERRAT-2009} for more details)
\begin{multline}\label{etr}
   \left[-i\frac{\Omega}{c_0}+\bm{u}\cdot\bm{\nabla}_{\bm{r}}+\int
      \mu_e\left(\delta-k_0\bm{u}\cdot\bm{v},\Omega\right)g\left(\bm{v}\right)\ud\bm{v}\right]
\\
   \times I\left(\bm{u},\bm{r},\delta,\Omega\right)=\frac{1}{4\pi}\int\mu_s\left(\delta-k_0\bm{u}\cdot\bm{v},\Omega\right)
\\
   \times I\left(\bm{u}',\bm{r},\delta+k_0\left(\bm{u}'-\bm{u}\right)\cdot\bm{v},\Omega\right)g\left(\bm{v}\right)\ud\bm{u}'\ud\bm{v}
\end{multline}
where $I$ is the specific intensity (local radiative flux at position $\bm{r}$, direction $\bm{u}$, frequency $\delta$
and time $t$, $\Omega$ being the Fourier variable for time). $g\left(\bm{v}\right)$ is the atomic velocity distribution
with mean-square $\vbar$ which writes
\begin{equation}
   g\left(\bm{v}\right)=\frac{1}{\left[\sqrt{2\pi}\vbar\right]^3}\exp\left[-\frac{\bm{v}^2}{2\vbar^2}\right].
\end{equation}
Equation~(\ref{etr}) is close to the standard Radiative Transfer Equation (RTE)~\cite{CHANDRASEKHAR-1950} except that
temporal convolution products and frequency shifts are present to take into account atomic resonances and Doppler
effects (finite temperature) respectively. The extinction and scattering coefficients are given by
\begin{align}\label{coeffs}
   \mu_e\left(\delta,\Omega\right) &
      =-\frac{i\rho k_0}{2}\left[\bm{\alpha}\left(\delta+\frac{\Omega}{2}\right)-\bm{\alpha}^*\left(\delta-\frac{\Omega}{2}\right)\right],
\\
   \mu_s\left(\delta,\Omega\right) &
      =\frac{\rho k_0^4}{4\pi}\bm{\alpha}\left(\delta+\frac{\Omega}{2}\right)\bm{\alpha}^*\left(\delta-\frac{\Omega}{2}\right)
\end{align}
where $\rho$ is the density of atoms. These expressions ensure energy conservation (the optical theorem is fulfilled).
The domain of interest was previously cold atoms where $k_0\vbar/\Gamma\ll 1$~\cite{PIERRAT-2009}.  In the case of hot
atomic vapors, the Doppler shift is larger than the natural linewidth.  Thus we have to take the limit of
Eq.~(\ref{etr}) when $k_0\vbar/\Gamma\gg 1$. In practice, we also perform a long time approximation by considering
$\Omega\ll\delta,\Gamma,k_0\vbar$. Using a Taylor expansion, we obtain the following expressions for the extinction and
scattering coefficients respectively:
\begin{align}\label{coeffs_approx}
   \mu_e\left(\delta-k_0\bm{u}\cdot\bm{v},\Omega\right)
      & =-\frac{i\rho k_0}{2}\left(1-\frac{i\Omega}{\Gamma}\right)
\\\nonumber
      & \times\left[\bm{\alpha}\left(\delta-k_0\bm{u}\cdot\bm{v}\right)-\bm{\alpha}^*\left(\delta-k_0\bm{u}\cdot\bm{v}\right)\right],
\\
   \mu_s\left(\delta-k_0\bm{u}\cdot\bm{v},\Omega\right)
      & =\frac{\rho k_0^4}{4\pi}\bm{\alpha}\left(\delta-k_0\bm{u}\cdot\bm{v}\right)\bm{\alpha}^*\left(\delta-k_0\bm{u}\cdot\bm{v}\right).
\end{align}
Moreover, the high temperature approximation allows us to approximate the integrations over the velocity. Indeed, the
polarizability becomes
\begin{multline}
   \bm{\alpha}\left(\delta-k_0\bm{u}\cdot\bm{v}\right) = -\frac{2\pi\Gamma}{k_0^3}\frac{1}{\delta-k_0\bm{u}\cdot\bm{v}+i\Gamma/2}
\\
   \approx -\frac{2\pi\Gamma}{k_0^3}\left\{
         \operatorname{VP}\left[\frac{1}{\delta-k_0\bm{u}\cdot\bm{v}}\right]
         -i\pi\bm{\delta}\left[\delta-k_0\bm{u}\cdot\bm{v}\right]
      \right\}
\end{multline}
where $\operatorname{VP}$ stands for the principal value operator and $\bm{\delta}$ is the Dirac delta function. Using
this expression, we find a simplified expression for the extinction part given by
\begin{multline}\label{extinction_approx}
   \mu_{\ell}\left(\delta\right)=\int\mu_e\left(\delta-k_0\bm{u}\cdot\bm{v},\Omega\right)g\left(\bm{v}\right)\ud\bm{v}
\\
      =\frac{\Gamma}{2\ell_0k_0\vbar}\sqrt{\frac{\pi}{2}}\exp\left[-\frac{\delta^2}{2k_0^2\vbar^2}\right]
\end{multline}
where $\ell_0$ stands for the scattering mean-free path for pinned atoms at resonance [i.e.\@
$\ell_0^{-1}=\mu_{s,e}\left(0,0\right)$].  In the following, we will essentially refer to a slab geometry. Denoting the
system thickness by $L$, it is possible to define the optical thickness by
\begin{equation}
   b\left(\delta\right)=\mu_{\ell}\left(\delta\right)L=\frac{L}{\ell\left(\delta\right)}
\end{equation}
where $\ell\left(\delta\right)$ is the scattering mean-free path at a given frequency.  Regarding the case of the
scattering part, an integral should be added over the frequency to simplify the one performed over the velocity.  We
define the phase function $p$ by
\begin{multline}\label{scattering_approx}
   \mu_{\ell}\left(\delta\right)p\left(\delta,\delta',\bm{u},\bm{u}'\right)=\int\mu_s\left(\delta-k_0\bm{u}\cdot\bm{v},\Omega\right)
\\
      \times\bm{\delta}\left[\delta'-\delta-k_0\left(\bm{u}'-\bm{u}\right)\cdot\bm{v}\right]g\left(\bm{v}\right)\ud\bm{v}
\end{multline}
and using the same approximation as for the extinction part, we obtain
\begin{multline}\label{phase_function}
   p\left(\delta,\delta',\bm{u},\bm{u}'\right)
      =\frac{1}{\sqrt{2\pi}k_0\vbar\sqrt{1-\left(\bm{u}\cdot\bm{u}'\right)^2}}
\\
      \times\exp\left[-\frac{\left(\delta'-\delta\bm{u}\cdot\bm{u}'\right)^2}
         {2k_0^2\vbar^2\left(1-\left(\bm{u}\cdot\bm{u}'\right)^2\right)}\right].
\end{multline}
The phase function describes the frequency and direction redistributions at each scattering event. This expression shows
that frequency redistribution is not complete, the scattered frequency being still related to the incident one through
the scattering angle given by $\bm{u}\cdot\bm{u}'$. On the other hand, the redistribution in direction is purely
isotropic because of the point character of the scatterers (atoms). For this reason, the scattering mean-free path
$\ell\left(\delta\right)$ and the transport mean-free path $\ell^*\left(\delta\right)$ are equal. Finally, we end-up
with the following transport equation
\begin{multline}\label{etr_levy}
   \left[\bm{u}\cdot\bm{\nabla}_{\bm{r}}+\left(1+\frac{1}{\Gamma}\frac{\partial}{\partial t}\right)\mu_{\ell}\left(\delta\right)\right]
   I\left(\bm{u},\bm{r},\delta,t\right)
\\
   =\frac{\mu_{\ell}\left(\delta\right)}{4\pi}\int p\left(\delta,\delta',\bm{u},\bm{u}'\right)
      I\left(\bm{u}',\bm{r},\delta',t\right)\ud\bm{u}'\ud\delta'.
\end{multline}
The transport equation effectively couples an equation of radiative transfer, an energy balance formulated in the
specific intensity, to a rate equation for the density of excited states. It is thus a generalization of the
Holstein-Bibermann integro-differential equation~\cite{HOLSTEIN-1947}. It is an inhomogeneous Fredholm equation of the
second kind and the kernel $p\left(\delta,\delta',\bm{u},\bm{u}'\right)$ of the integral transform quantifies the
qualitative signature of superdiffusion:~This kernel is long-ranged, making transport non-local and thus invalidating
further simplification which could end up in a diffusion-like formulation.  In its present format, it also captures
memory effects due to inelastic scattering. One should also note that a similar equation to Eq.~(\ref{etr_levy}) was
derived previously in the steady-state regime in the context of planetary nebula~\cite{UNNO-1952}. The resolution of
Eq.~(\ref{etr_levy}) is possible numerically using an exact Monte Carlo scheme as detailed in
appendix~\ref{monte_carlo}.

\subsection{Step-size distribution and L\'evy flights}\label{theory-levy-assymptotic}

To go further with an analytical approach and show that L\'evy flights are encoded in this equation, we need to perform
the approximation $\bm{u}\cdot\bm{u}'=0$. This assumption is true on average (isotropic scattering) and holds only if a
large number of scattering events are encountered by photons before leaving the system. We have checked numerically the
validity of this assumption by Monte Carlo simulations and found no relevant deviations in the case of large optical
thicknesses as those considered in the following. It implies complete frequency redistribution and the kernel of
the integral transform is absent of any memory from past history. From Eq.~(\ref{etr_levy}), we obtain
\begin{multline}\label{etr_levy_uu}
   \left[\bm{u}\cdot\bm{\nabla}_{\bm{r}}+\left(1+\frac{1}{\Gamma}\frac{\partial}{\partial t}\right)
      \mu_{\ell}\left(\delta\right)\right]I\left(\bm{u},\bm{r},\delta,t\right)
\\
   =\frac{\mu_{\ell}\left(\delta\right)}{4\pi} \int p\left(\delta'\right)I\left(\bm{u}',\bm{r},\delta',t\right)\ud\bm{u}'\ud\delta'
\end{multline}
where
\begin{equation}
   p\left(\delta'\right)=\frac{1}{\sqrt{2\pi}k_0\vbar}\exp\left[-\frac{\delta'^2}{2k_0^2\vbar^2}\right]
\end{equation}
This equation allows us to derive the expression of the step-size distribution and to compute analytically its
asymptotic behavior given by
\begin{equation}\label{ps}
   P\left(x\right)=\int \mu_{\ell}\left(\delta\right)\exp\left[-\mu_{\ell}\left(\delta\right) x\right]
      p\left(\delta\right)\ud\delta\propto 1/x^{\left(\alpha+1\right)}
\end{equation}
when $x\to\infty$ with $\alpha=1$. In this limit the average and the variance of the scattering length are not defined,
leading to a diverging diffusion coefficient, confirming the presence of L\'evy flights induced by large Doppler shifts
(high temperature) in conjunction with sharp resonances (high quality factor of the atomic transition).

From a random walk point of view, it has also been shown that this asymptotic scaling $P(x)\propto 1/x^{\alpha+1}$ of
the step-size distribution of independent steps implies a power law behavior for the steady-state total diffuse
transmission $T_{\textrm{diff}}$ through a slab of thickness $L$. More precisely, it is given by
$T_{\textrm{diff}}\propto L^{-\alpha/2}$ at large optical thicknesses~\cite{MARSHAK-1997,WIERSMA-2008-1}.  In our case,
this scaling is given by $1/\sqrt{L}$ and is very different from the classical diffusion case where the total diffuse
transmission is proportional to $1/L$. This has been check numerically using the Monte Carlo algorithm and an excellent
agreement is obtained between analytics and numerics.

\begin{figure}[!htbf]
   \centering
   \includegraphics[width=0.8\linewidth]{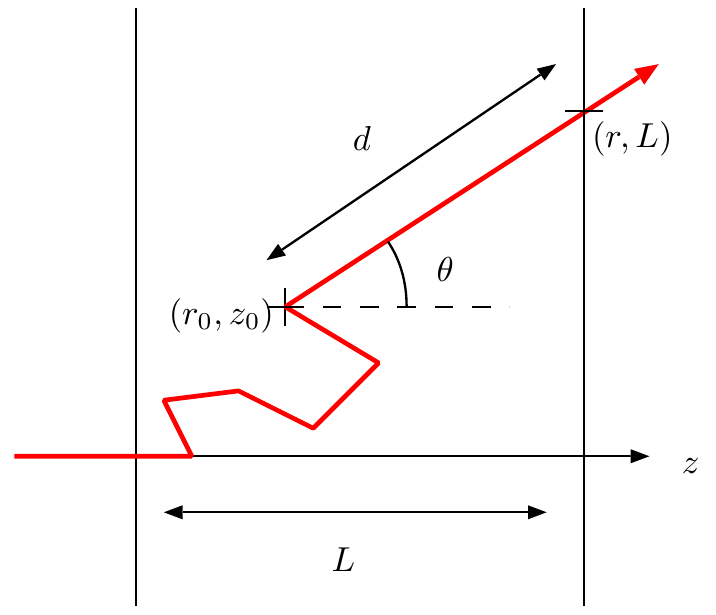}
   \caption{Notations used in a slab geometry to derive the radial dependance of the transmitted flux in the case of L\'evy flights.}
   \label{radial}
\end{figure}

Still in the case of a slab geometry, we have exploited Eq.~(\ref{ps}) further to show analytically that the transmitted
intensity profile $T_{\textrm{diff}}(r)$ has power law tails. We assume that the last step is predominant compared to
all others steps if photons escape the system far from the axis (i.e.\@ $r\gg L$ if $r$ is the radial distance). Let's
assume that the position of the last scattering event before exit is $(r_0,z_0)$ as shown on Fig.~\ref{radial}. Then,
the transmitted intensity escaping the system in the range $[\mu,\mu+\ud\mu]$ is
\begin{align}
   \ud T_{\textrm{diff}} & =\int_d^{+\infty}P\left(x\right)\ud x\,\ud\mu
\end{align}
where $\mu=\cos\theta=\left(L-z_0\right)/d$ and $d^2=\left(r-r_0\right)^2+\left(L-z_0\right)^2$. Considering large $r$
compared to the size of the system, we obtain
\begin{equation}
   \ud T_{\textrm{diff}}=T_{\textrm{diff}}\left(r\right)r\ud r\propto \frac{r\ud r}{r^{\alpha+3}}
\end{equation}
proving that
\begin{equation}
   \label{Ir}
   T_{\textrm{diff}}(r)\propto r^{-3-\alpha}\propto 1/r^4.
\end{equation}
The tails of the transmitted intensity profile in multiple scattering conditions can thus be used to measure the
microscopic L\'evy exponent $\alpha$. We stress that the possibility to extract the L\'evy exponent from one
single transmission profile has an important advantage, as one does not need to realize samples with different size or
opacity, already often difficult in laboratory environment, but impossible to realize in the astrophysical context. As
for the total diffuse transmission, we have checked the validity of Eq.~(\ref{Ir}) by the Monte Carlo algorithm. The
agreement between the theory and the numerics is very good as shown in Fig.~\ref{fig5}\,(b) where the wings of the
profile develop a well pronounced power law, in contrast to the wings obtained in the classical diffusive limit, where
this profile has a quasi exponential tail as shown in section~\ref{classical-diffusion}.

\subsection{Long time behavior and L\'evy flights}

A L\'evy flights signature can also be derived by looking at the dispersion relation of Eq.~(\ref{etr_levy_z}). To do
so, we still consider a slab geometry of thickness $L$. If the system is illuminating from the left by a plane wave at
normal incidence, the problem is invariant by rotation on the azimuthal angle $\varphi$ and by translation along $x$ and
$y$. By integrating Eq.~(\ref{etr_levy}) over $\varphi$, we obtain
\begin{multline}\label{etr_levy_z}
   \left[\mu\frac{\partial}{\partial z}+\left(1+\frac{1}{\Gamma}\frac{\partial}{\partial t}\right)
      \mu_{\ell}\left(\delta\right)\right]I\left(\mu,z,\delta,t\right)
\\
   =\frac{\mu_{\ell}\left(\delta\right)}{2} \int p\left(\delta'\right)I\left(\mu',z,\delta',t\right)\ud\mu'\ud\delta'.
\end{multline}
We now perform a modal expansion of the specific intensity by taking its spatial Fourier transform combined with
temporal eigenmodes~\cite{CASE-1967,PIERRAT-2006,PIERRAT-2009}) which leads to
\begin{equation}
   I\left(\mu,z,\delta,t\right)=g_{ks}\left(\mu,\delta\right)\exp\left[ikz+s\left(k\right)t\right].
\end{equation}
$g_{ks}$ can be interpreted as the eigenvector associated to the eigenvalue $s\left(k\right)$. Injecting this expression
into Eq.~(\ref{etr_levy_z}) and defining a renormalized eigenvector by
$G_{ks}\left(\mu,\delta\right)=p\left(\delta\right)g_{ks}\left(\mu,\delta\right)$, we obtain the dispersion relation of
the transport equation given by
\begin{equation}
   \int\frac{\mu_{\ell}\left(\delta\right)p\left(\delta\right)}{k}
      \arctan\left[\frac{k}{\left(1+s/\Gamma\right)\mu_{\ell}\left(\delta\right)}\right]\ud\delta=1.
\end{equation}
To obtain an explicit relation on $s$, we define $K=k/\left(1+s/\Gamma\right)$. The predominant spatial mode is such
that $k=\pi/L$~\cite{PIERRAT-2006} where $L$ is the thickness of the slab. This implies that when $L\to\infty$, $k\to 0$
and $s\to 0$. This leads to $K\approx k$ and finally we obtain
\begin{equation}
   \frac{s}{\Gamma}\approx\int_{-\infty}^{+\infty}\frac{\mu_{\ell}\left(\delta\right)
      p\left(\delta\right)}{k}\arctan\left[\frac{k}{\mu_{\ell}\left(\delta\right)}\right]\ud\delta-1
\end{equation}
when $k\to 0$. This equation gives the temporal decay rate of the transmitted flux for long times and as a function of
the system size:
\begin{equation}
   T_{\textrm{diff}}\left(t\right)\propto \exp\left[s\left(\frac{\pi}{L}\right) t\right].
\end{equation}
This behavior could be observed in time resolved experiments which has already been done for quenched
systems~\cite{ARXIV-SAVO-2014}.  By taking the limit $k\to 0$ (large system sizes), we find $s\propto k^{\alpha}$ with
$\alpha=1$ leading to an anomalous diffusion equation with a spatial derivative $\Delta^{\alpha/2}$ which confirms again
the presence of  L\'evy flights~\cite{KLAFTER-1999}.

Note that a link can be made easily between $s\left(k\right)$ and the step-size distribution $P\left(x\right)$. We find
\begin{equation}
   \frac{s\left(k\right)}{\Gamma}=\int_{0}^{+\infty}\frac{\sin\left(kx\right)}{kx}P\left(x\right)\ud x-1.
\end{equation}
This expression shows that $ks$ is given by the Fourier transform of $P/x$ which implies that if $P\left(x\right)\propto
x^{-\alpha-1}$ when $x\to\infty$ then $s\left(k\right)\propto k^{\alpha}$ when $k\to 0$.

\subsection{Standard diffusion radial profiles}\label{classical-diffusion}

In~\ref{theory-levy-assymptotic} the focus was on the asymptotic of the spatial distributions for L\'evy flights. This
asymptotic gives its characteristic signature. We found that the single-step fundamental L\'evy $\alpha$
parameter [$P\left(x\right)\propto 1/x^{\alpha+1}$] survives in the multiple-step regime for a slab
geometry in the form of a power law in the radial transmission profile [$T_{\textrm{diff}}(r)\propto r^{-3-\alpha}=
1/r^4$]. Our main goal is to characterize unambiguously the L\'evy regime, emphasising its qualitative difference from
standard diffusion. We therefore tried to address the following question:~what is the asymptotic in the multiple
scattering regime in the spatial radial profile for standard diffusion? Is it exponential (possibly mimicking a single
step step-size distribution); or alternatively, is it Gaussian (mimicking the Gaussian propagator in infinite media)?
In trying to address this question we derived an analytical representation of the radial profiles in the slab for both
transmission and reflection. This indeed gives a quasi-exponential asymptotic for standard diffusion, thus emphasizing
an important physical insight. The qualitative nature of the single-step distribution still survives in the multi
scattering regime: a L\'{e}vy signature can be inferred from a power law in the radial profile and a exponential radial
dependence is the asymptotic signature of standard diffusion. The radial profiles in an infinite slab, for isotropic
scattering and no absorption can be represented as:
\begin{align}\nonumber
   R_{\textrm{diff}}\left(r\right) =\, & \frac{\pi}{4}
      \left(\frac{1}{L_{e}}\right)^{2}i\left\{
         \sum_{n=1}^{+\infty}n\,\cos\left(n\pi\frac{z_{e}}{L_{e}}\right)
      \right.
\\\label{diff_eq_solution R} & \times
      \left.\vphantom{\sum_{n=1}^{+\infty}}
         \sin\left(n\pi\frac{\ell^*+z_{e}}{L_{e}}\right)
         \operatorname{H}_{0}^{\left(1\right)}\left(in\pi\frac{r}{L_{e}}\right)
      \right\},
\\\nonumber
   T_{\textrm{diff}}\left(r\right) =\, & \frac{\pi}{4}
      \left(\frac{1}{L_{e}}\right)^{2}i\left\{
         \sum_{n=1}^{+\infty}\left(-1\right)^{n+1}n\,\cos\left(n\pi\frac{z_{e}}{L_{e}}\right)
      \right.
\\\label{diff_eq_solution} & \times
      \left.\vphantom{\sum_{n=1}^{+\infty}}
         \sin\left(n\pi\frac{\ell^*+z_{e}}{L_{e}}\right)\operatorname{H}_{0}^{\left(1\right)}\left(in\pi\frac{r}{L_{e}}\right)
      \right\},
\end{align}
where $L$ is the slab width, $\ell^*$ the transport mean-free path, and $L_{e}=L+2z_{e}$ an effective width,
with~$z_{e}$ an extrapolation length added to take into account the boundary conditions.
$R_{\textrm{diff}}\left(r\right)$ and $T_{\textrm{diff}}\left(r\right)$ are the radial profiles in reflection and
transmission. For details of the derivation, see~appendix~\ref{appendix-diffusion}.

\begin{figure}[!htbf]
   \centering
   \includegraphics[width=\linewidth]{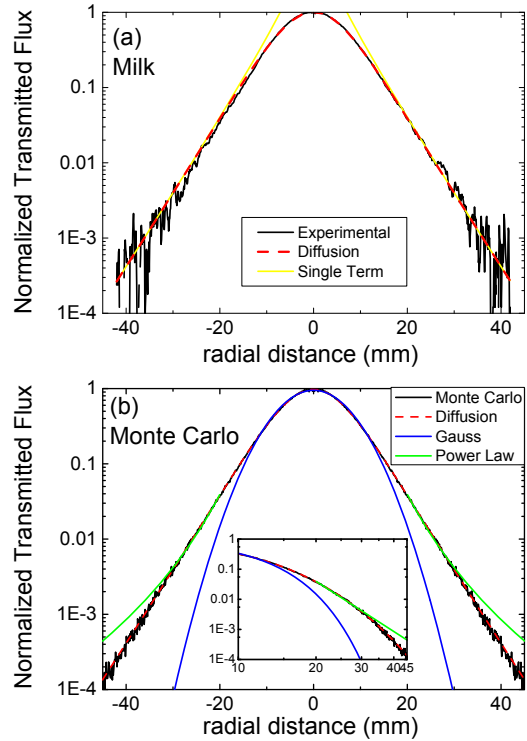}
   \caption{(Color online)~Transmission profiles (experimental, with milk, and for a Monte Carlo simulation) and
   corresponding fits with an analytical model for diffusion, Eq.~(\ref{diff_eq_solution}).  (a)~Experimental
   transmission profile ($\ell^*=\unit{0.05 \pm 1.5}{\milli\meter}$, $L=\unit{14.82 \pm 0.05}{\milli\meter}$).
   Also shown is the contribution of a single term, the first and slowest decaying of the series,
   the one responsible for the quasi-exponential asymptotic.  (b)~Monte Carlo (input parameters from the fit of the milk)
   for an infinite slab, fitted with the analytical model Eq.~(\ref{diff_eq_solution}) (three fitting parameters: 
   width, transport mean-free path and scale;~$\ell^*=\unit{0.09 \pm 2.5}{\milli\meter}$, 
   $L=\unit{14.7 \pm 3.5}{\milli\meter}$), with a Gaussian (two parameters: width and scale)
   and with a power law asymptotic law (two parameters: power law parameter and scale, plus a cut-off distance;
   here~$\unit{20}{\milli\meter}$). The diffusion model is indistinguishable from the Monte Carlo data (even in a log
   scale spanning four orders of magnitude).  The Gaussian and the power law are inadequate.}
   \label{fig2}
\end{figure}

The~$\operatorname{H}_{n}^{\left(1\right)}$ are the Hankel~$\operatorname{H}$ functions (Bessel functions of the third
kind). Two properties of the Hankel functions are important for this work: their divergence at the origin and their
asymptotic behaviour. We checked and, although each term in the previous series diverges by itself, the series is indeed
convergent at the origin (details are given in the appendix~\ref{appendix-diffusion}). More important is the asymptotic
behavior. The asymptotic valid for large distances from the center (please remember that it is exactly this asymptotic
behaviour that we sough) is easily obtained. The asymptotic for each term is~\cite{DUFFY-2001}:
\begin{equation}
   \label{diff_eq_asymptotic}
   i\, \operatorname{H}_{0}^{\left(1\right)}\left(in\pi\frac{r}{L_{e}}\right)
      \sim\frac{1}{\pi}\,\sqrt{\frac{2L_{e}}{n}}\, r^{-1/2}\,\exp\left(-n\pi\frac{r}{L_{e}}\right).
\end{equation}
For large distances, only the first term of the series survives and the asymptotic is approximately
exponential [see Fig.~\ref{fig2}(a)]. What emerges is an approximate
exponential asymptotic for both the transmitted and reflected radial profiles. This quasi-exponential asymptotic can be
interpreted as the signature of standard diffusion in the slab. This dependence is present in the analytical solution of
the diffusion equation, as well as in the solution of the radiative transfer equation, not restricted to the diffusion
approximation~\cite{LIEMERT-2012,MACHIDA-2010}. Although the exponential asymptotic is only approximate,
the~$r^{-1/2}$ dependence is weak and in all practical situations a log-lin plot gives an easy graphical criteria to
identify a diffusion-like asymptotic [see Fig.~\ref{fig2}].

We checked that the sum of the overall reflection and transmission intensities, obtained by integrating over angle from
Eqs.~(\ref{diff_eq_solution R}) and~(\ref{diff_eq_solution}), is one. This is of course the expected result based on physical
grounds:~since absorption was not considered, this trivially expresses energy conservation.

Fig.~\ref{fig2} shows fits of the transmitted radial profile with the analytical model of Eq.~(\ref{diff_eq_solution}),
for both an experimental result (milk) as well as a Monte Carlo simulation. The fits with the diffusion model are very
good (for the Monte Carlo, the fit is perfect, better than real life) and the agreement between the known slab width and
the one recovered by fit is also very good:~for milk, experimental~$L=\unit{16\pm 1}{\milli\meter}$, by
fit~$L=\unit{14.82 \pm 0.05}{\milli\meter}$; for Monte Carlo (with input parameters corresponding to the fitted ones,
obtained from milk), one recovers by fit~$L=\unit{14.7 \pm 3.5}{\milli\meter}$. Part~(a) puts in evidence the
quasi-asymptotic regime and further illustrates the relative contribution of the first term in the series, the one
giving the quasi-asymptotic, to the overall profile. Part~(b) shows Monte Carlo data, also fitted with a simple Gaussian 
and a Power Law model. We use the Monte Carlo data to illustrate the inability of the Gaussian to adequately describe 
the data (also true for the experimental data). Fig.~\ref{fig2} is important since
it shows perfect agreement with the diffusion prediction of Eq.~(\ref{diff_eq_solution}), but also because it shows that
the difference of a Gaussian to the analytical model is accessible in an actual experiment (our experimental setup
easily access two to four orders of magnitude; the standard practice of plotting the profile in lin scale makes the
disagreement with the Gaussian much less evident). It is also operational important since it dismiss any possible
artifacts in the Monte Carlo general scheme (the general framework in which the Monte carlo of the L\'evy flight, a far
from trivial code, is constructed). And illustrates a simple graphical procedure to diagnose the qualitative signature
of the data:~a log-lin plot will signal diffusion, a log-log plot a power law (see Fig.~\ref{fig5}, below).

In Fig.~\ref{fig2}, the confidence interval for the fitted values of the mean transport length in very large. This finding was
the motivation to make the Monte Carlo simulation using as input parameters the ones obtained from the fit of the milk data:~we
wanted to test the ability to recover meaningful values for the transport length. Although the agreement for $\ell^*$ between 
input and recovered by fit is satisfactory only when considering the corresponding confidence intervals, we found that in this 
well developed diffusion regime ($L/\ell^*=274$, for milk), Eq.~(\ref{diff_eq_solution}) is very insensitive to the
actual~$\ell^*$ value. Therefore, in this regime, the transmitted profile is not the best observable to access 
meaningfully~$\ell^*$ (tests not shown). This is the reason of a very broad confidence interval 
for the~$\ell^*$ values of course. Additional technicalities of the fitting algorithm and limitations are
given in appendix~\ref{appendix-diffusion}.

The quality of the fit and the good agreement of the Monte Carlo data and the diffusion model illustrate that, at least for 
aspect ratios not smaller than the ones used in this work (for milk the ratio of slab width to diameter is roughly~$10$), the
diffusion model assuming an infinite slab is an adequate description of an actual experiment (please remember that both 
analytical model and Monte Carlo correspond to an infinite slab; actual experiment is always finite).

An important physical insight can be stressed at this point. Altough the actual details of standard and anomalous
diffusion can be quite involved, their microscopic physical signature survives in the asymptotics of steady state in a
strongly multiple scattering regime: standard diffusion with an exponential single step distribution gives rise to
asymptoticaly exponential spatial profiles (and not to Gaussian profiles, as one might guess from the Gaussian
propagator in infinite media) while a single step power law L\'evy will reveal itself as a power law in the multiple
step spatial profile. This insight is important since the fundamental nature of diffusion can be investigated in
multiple scattering conditions, usually much easier to access than the corresponding single scattering counterparts.

\section{Experiment}\label{2nd_part}

\subsection{Setup description}

Experimentally, microscopic measurements of the step-size distribution have already been realized~\cite{KAISER-2009}
showing the existence of L\'evy flights of light in hot atomic vapors. In this article, we focus on the macroscopic
evidence of L\'evy flights by recovering the L\'evy parameter $\alpha$ using a stationary experiment under multiple
scattering conditions. Our experiment is really close to ideal Gedanken experiment. We consider a flat cell (Petri dish
shape made of pirex) of diameter $\unit{10}{\centi\meter}$, external thickness $\unit{11}{\milli\meter}$ and internal
thickness $L=\unit{5}{\milli\meter}$. It contains a natural mixture of 85 and 87 rubidium vapor. This cell is
illuminated by a distributed feedback (DFB) laser whose
spectral width is on the order of \unit{2}{MHz} and the maximum of power before cell can reach several milliwatts. Not
to saturate atoms and CCD camera we use only few dozen microwatts after several attenuators.  One part of the incident
laser is used for rubidium spectroscopy and for locking on the $F=2\to F'=2\relbar F'=4$ crossover of the D2 line of
\textsuperscript{85}Rb. The laser has a piedestal \unit{40}{dB} below the laser peak but with some nanometers of width.
This piedestal can be a problem since, even if it is out of resonance of rubidium and does not scatter on rubidium, a
part scatters on the sides of cell. The main part of the laser beam is injected in a monomode fiber whose output shines
the center of the cell. The laser is collimated with a waist of $w\approx\unit{1.3}{\milli\meter}$. The cell is placed on
an oven whose temperature can be controlled homogeneously between $\unit{40}{\celsius}$ to $\unit{170}{\celsius}$,
therefore tuning the density of atoms $\rho$ between around $\unit{6\times\power{10}{15}}{at/m^3}$ and
$\unit{2\times\power{10}{20}}{at/m^3}$.  The laser source is one meter far from oven to avoid temperature control
problems.

Experimentally, the optical thickness $b\left(\omega\right)$ and the extinction coefficient
$\mu_{\ell}\left(\omega\right)$ are not the most convenient quantities to characterize the optical response of the cell
in particular because of their dependences on the frequency. We prefer to use in the following the opacity $O$ and the
scattering cross section $\sigma_T\left(\omega\right)$. The opacity is defined by
\begin{equation}
   O=\frac{L}{\ell_0}.
\end{equation}
As $\ell_0$ depends on the density $\rho$, it is possible to tune in a well controlled manner the opacity by just
changing the temperature. In particular, the range of temperature considered in the experiment corresponds to a change
in opacity $O$ from $\power{10}{2}$ to $\power{10}{5}$.  The opacity is thus a good indicator of an effective cell
thickness and the scaling laws for $O$ and $L$ are identical. Using this definition, the optical thickness becomes
\begin{equation}
   b\left(\omega\right)=\frac{L}{\ell\left(\omega\right)}=O\frac{\ell_0}{\ell\left(\omega\right)}
      =O\frac{\sigma_T\left(\omega\right)}{\sigma_0}
\end{equation}
where $\sigma_0$ is the scattering cross section of a pinned atom at resonance. $\sigma_T\left(\omega\right)$ is given by
\begin{equation}
   \sigma_T\left(\omega\right)=\int\sigma\left(\omega-\bm{k}\cdot \bm{v}\right)g\left(\bm{v}\right)\ud\bm{v}
\end{equation}
with the scattering cross section for atoms at rest given by
\begin{equation}
   \sigma\left(\omega\right)=\sum_{i}R_i\pi_i\frac{\sigma_0}{1+4\delta_i^2},
\end{equation}
$R_i$ being the branching ratios of each hyperfine line $i$ of the D2 line of \textsuperscript{85}Rb and
\textsuperscript{87}Rb. The detunings are given by $\delta_i=\left(\omega-\omega_{0,i}\right)/\Gamma$ where
$\omega_{0,i}$ are resonance frequencies and $\Gamma$ the linewidth.

\subsection{Ballistic transmission measurement}

\begin{figure}[!htbf]
   \centering
   \includegraphics[width=\linewidth]{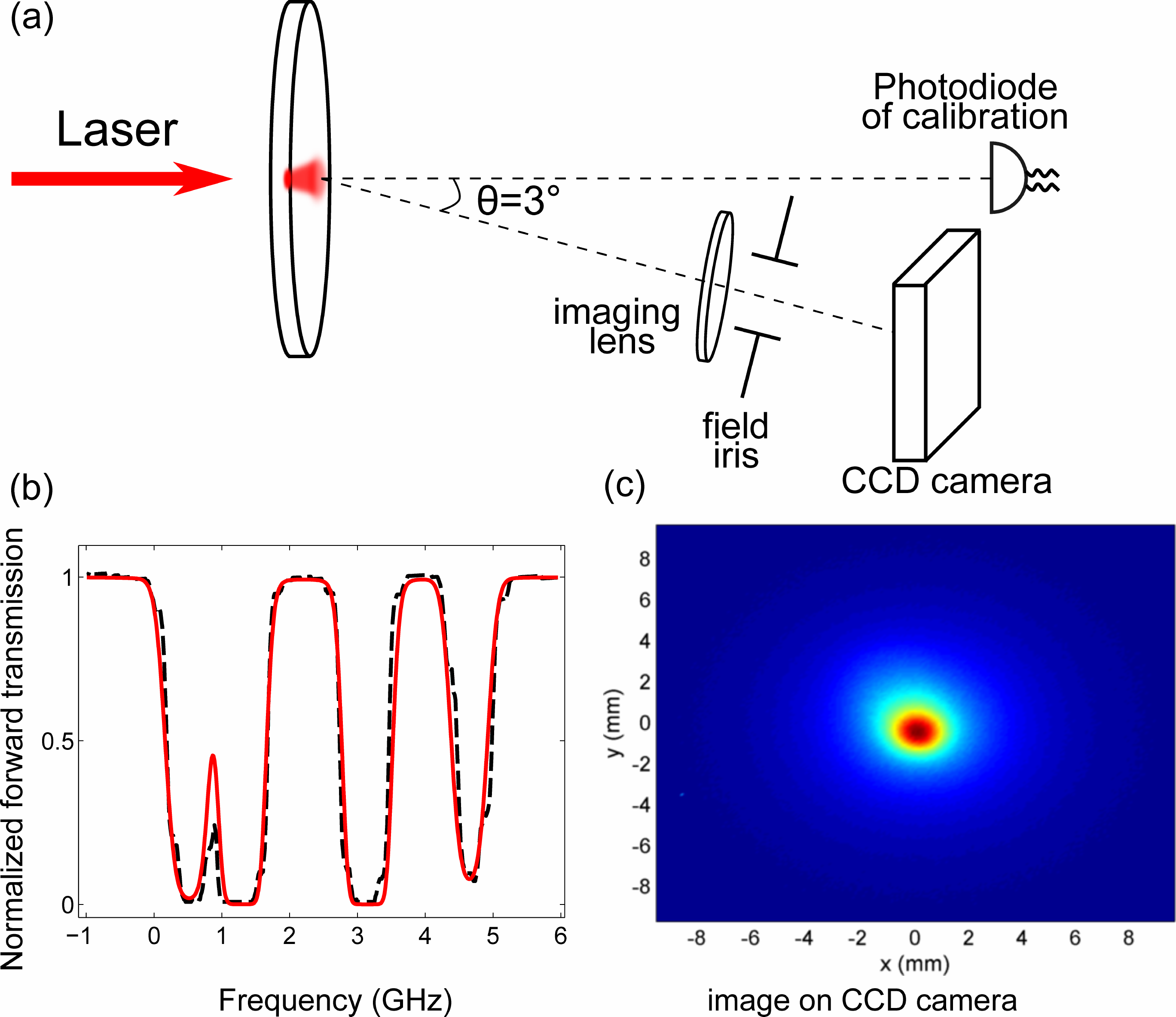}
   \caption{(Color online) (a) Experimental setup: A laser beam is incident on a flat cylindrical cell filled with a
   rubidium vapor. The transmitted ballistic light spectrum is measured via a photodiode aligned with the laser beam.
   The scattered light coming out of the slab is imaged on the CCD camera misaligned by an angle of $\unit{3}{\degree}$
   with the laser. (b) Signal of the photodiode: The transmission of the laser scanned in frequency throughout the
   rubidium cell shows the D2 hyperfine absorption of \textsuperscript{85}Rb and \textsuperscript{87}Rb.  An ab-initio
   fit of this signal is used to deduce the opacity $O$. Even if the system is strongly scattering, a coherent
   transmission measurement can still be done especially out of resonance where the optical thickness is weaker. (c)
   Image of the radial profile observed on the CCD camera.}
	\label{fig3}
\end{figure}

Experimentally, the opacity $O$ is estimated by measuring the coherent transmission
$T_{\textrm{coh}}\left(\omega\right)$ of a laser beam through the cell using a photodiode [see Fig.~\ref{fig3}\,(a)]. It
is given by
\begin{equation}
   T_{\textrm{coh}}\left(\omega\right)
      =\exp\left[-\rho L\sigma_T\left(\omega\right)\right]=\exp\left[-O\frac{\sigma_T\left(\omega\right)}{\sigma_0}\right].
\end{equation}
The incident frequency is tuned to scan all D2 lines of two rubidium isotopes [see Fig.~\ref{fig3}\,(b)]. In our range
of temperature, the opacity scales as $O\sim 200\log\left(T\right)$ which ensures the multiple scattering regime. The
spectral broadening is proportional to the square root of the temperature while the vapor pressure depends exponentially
on the temperature ensuring that the Doppler effect is almost not affected by the temperature.

\subsection{Diffuse transmission measurement}

\begin{figure}[!htbf]
   \centering
   \includegraphics[width=0.8\linewidth]{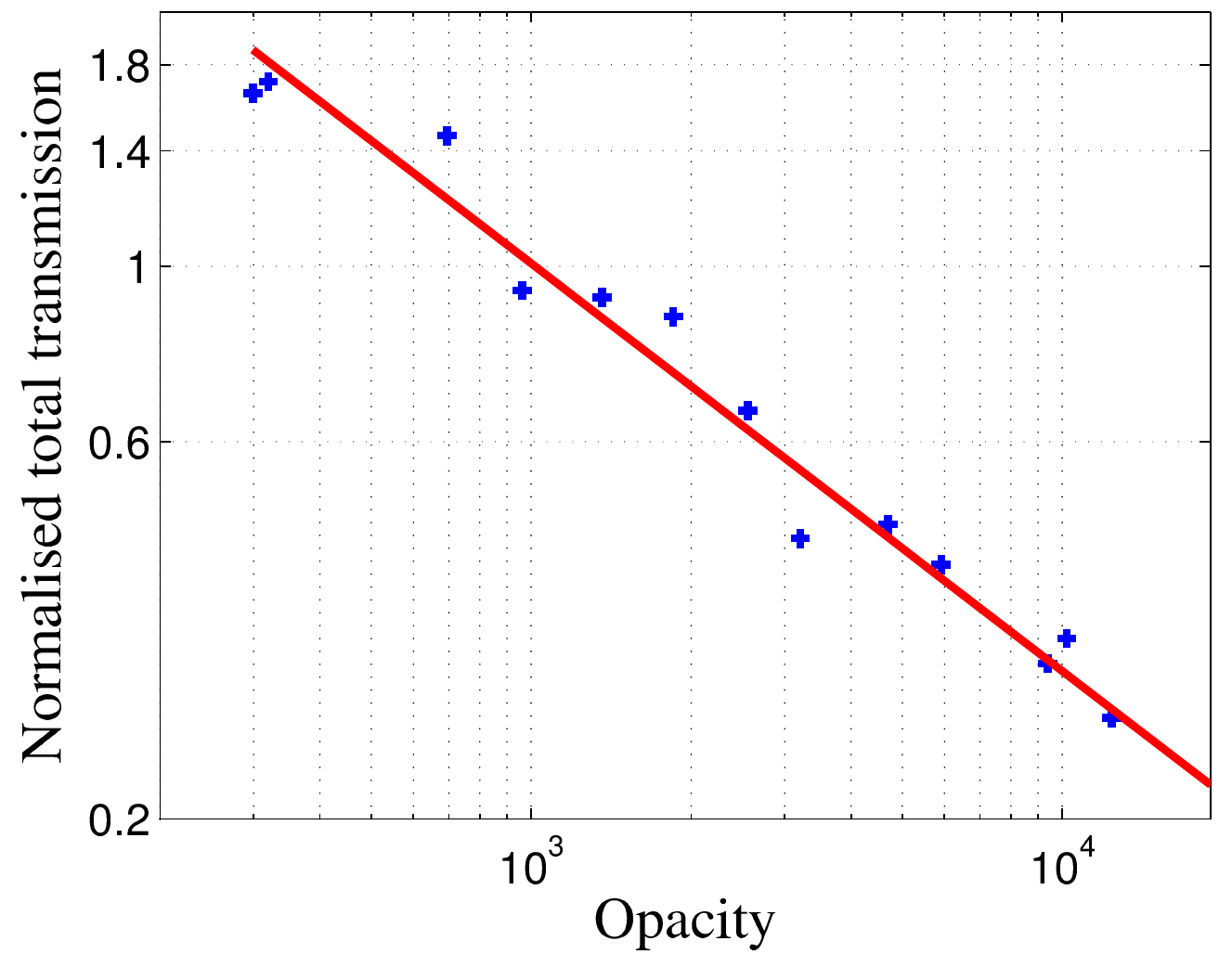}
   \caption{(Color online) Total diffuse transission in log scale as a function of the opacity. Blue crosses:
   Experimental data.  Each point corresponds to the sum of all pixels of a CCD image for a given opacity. Black solid
   line: fit of the experimental data. The power law obtained is $T_{\textrm{diff}}\propto O^{-0.516 \pm 0.024}$.}
   \label{fig4}
\end{figure}

We now turn into the stationary diffuse transmission measurement. For long term stability and an absolute frequency
reference, we lock the laser on the $F=2\to F'=2\relbar F'=4$ crossover of the D2 line of \textsuperscript{85}Rb. 

\begin{figure}[!htbf]
   \centering
   \includegraphics[width=0.7\linewidth]{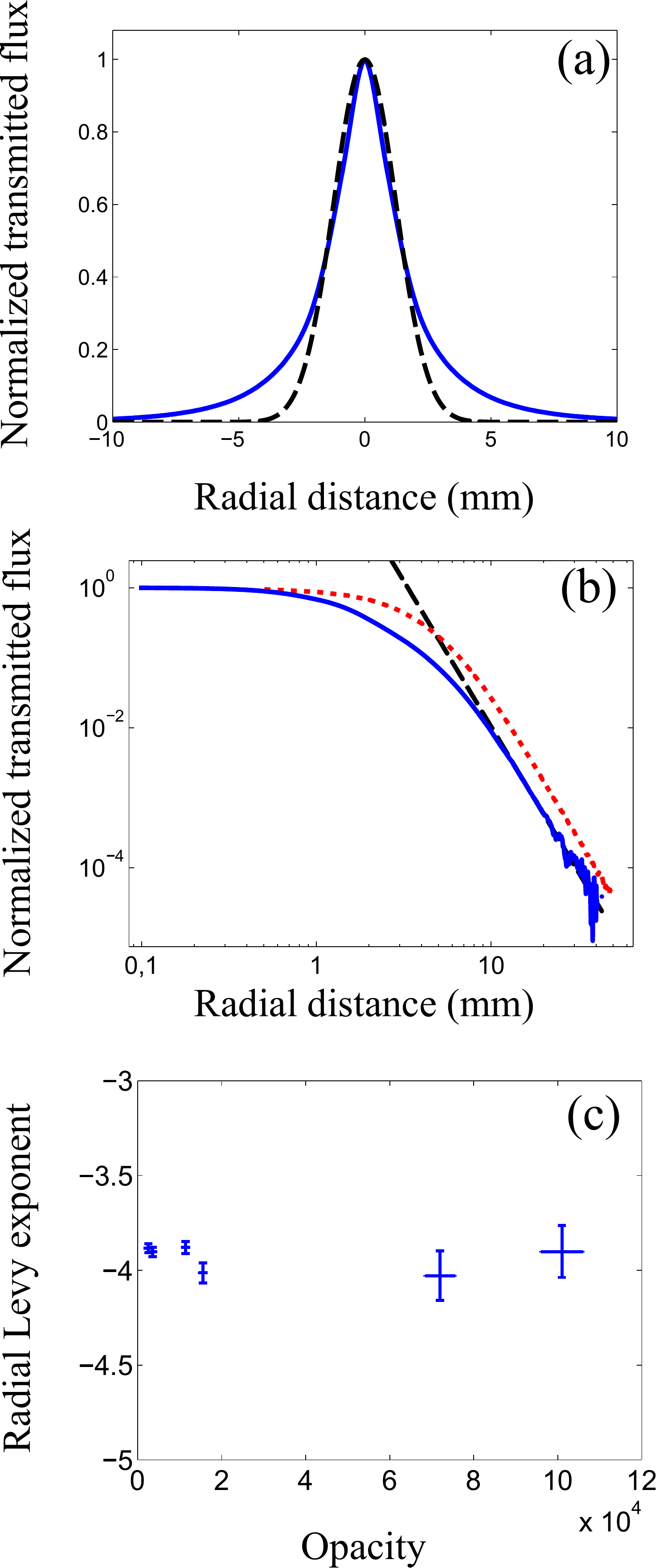}
   \caption{(Color online) (a) Blue solid line: Experimental radial profile (in linear scale) of the scattered light
   transmitted through the cell. Black dashed line: Gaussian profile with the same width at half maximum as the
   experimental profile. (b) Blue solid line: Experimental radial profile (in log scale) of the scattered light
   transmitted through the cell. Dotted red line: Numerical prediction from a Monte Carlo simulation of the transport
   equation. Black dashed line: fit of the experimental data in the range $\unit{[1.1, 2.1]}{\centi\meter}$.  The power
   law obtained is $T_{\textrm{diff}}(r)\propto r^{-4.03 \pm 0.15}$. The temperature is $T=\unit{114}{\celsius}$
   corresponding to an opacity of $O=\unit{1.15} \times {\power{10}{4}}$ and an optical thickness of $b=540$. (c) Power law
   exponent of the radial profile tail of the transmitted intensity $T_{\textrm{diff}}(r)$ as a function of the opacity
   $O$ tuned by varying the temperature. This exponent is close to $-4$ for a wide range of opacity showing the
   robustness of the L\'evy character of the system. Vertical error bars correspond to standard deviation of the fit to
   extract the power law of $T_{\textrm{diff}}(r)$. Error bars increase for large opacities due to the decrease of the
   relevant signal.}
   \label{fig5}
\end{figure}

Due to a residual background of amplified spontaneous emission (ASE) of our DFB laser, scattering of this spectrally
broad pedestal on the facets of our rubidium cell give rise to an additional offset signal. As we focus on the power law
tails in the image, the center of the image (where such background scattering is present) does not affect the result of
our analysis. We however noted that when using an imaging system with a \unit{25}{\milli\meter} diameter lens, the depth
of field (DOF) was small enough to blurr the image of the scattered component of the ASE on the input facet of our cell.
Indeed, a small DOF results in a diffractive type power tail in the detected image of light scattered from an object
beyond the DOF. Even though the power in the pedestal of our DFB laser only represents about \unit{0.4}{\%} of the power
of the laser, this can limit the possibility to detect the relevant power law tails under consideration. This effect
would be a crucial limititation to the light scattered by the atoms at large distances. One solution would be to use a
laser with a reduced ASE background. In our setup, we solved this problem by decreasing the diameter of an iris placed
in front of the imaging lens from \unit{25}{\milli\meter} to \unit{10}{\milli\meter}. We thus deliberately increased the
DOF beyond the length of our rubidium cell. Light scattered on the windows of both input and output facets of the cell
are thus only visible at the center of the image at the CCD, not contributing to the wings of the radial profile, where
we extract the L\'evy flight information from the light scattered by the atoms.  We have checked that using an iris
diameter of \unit{10}{\milli\meter}, the radial profile measured by the CCD camera, with the laser tuned out of the
atomic resonance line, decreases by more than 4 orders of magnitude at only \unit{2}{\milli\meter} whereas with a fully
open iris, the same reduction is obly obtained at \unit{10}{\milli\meter}. As the image of the relevant output spot is
small compare to CCD camera, we use the pixels in peripheria to define the background signal to subtract.  Increased
signal to noise ratio is obtained by averaging with an interpolation method over $36$ cuts across the center of the
image (each rotated by \unit{10}{\degree}). Subsequent smoothing of the data further improves the signal to noise ratio
shown in Fig.~\ref{fig5}. 

In Fig.~\ref{fig4} we plot the total transmission incoming on all pixels of the CCD camera as a function of the
opacity. This does not correspond exactly to the total transmitted energy outgoing the system because of the small
numerical aperture considered.  Nevertheless, we have checked numerically using the Monte Carlo simulation that this
does not affect the scaling law.  We clearly see that the total transmission $T_{\textrm{diff}}$ does not scale as Ohm's
law prediction (i.e.\@ $T_{\textrm{diff}} \propto O^{-1}$) but has a superdiffusive behavior (i.e.\@ $T_{\textrm{diff}}
\propto O^{-0.516}$), in excellent agreement with the prediction at large opacities $T_{\textrm{diff}}\propto
O^{-\alpha/2}$ for $\alpha\approx1$~\cite{MARSHAK-1997,WIERSMA-2008-1}.  Considering the single step size distribution
measured in \cite{KAISER-2009} this exponent is in excellent agreement with the expectation for the total diffuse
transmission for a L\'evy walk of independent steps and different from random walk in quenched
disorder~\cite{BEENAKKER-2012}.

Exploiting the excellent signal to noise ratio of our angular averaged CCD images, we also plot the radial profile of
the transmitted intensity $T_{\textrm{diff}}(r)$ in linear [see Fig.~\ref{fig5}\,(a)] and logarithmic [see
Fig.~\ref{fig5}\,(b)] scales, highlightening the large dynamic range of our signal. We clearly see that the general
shape is far from a gaussian profile.  The tail clearly exhibits a power law: $T_{\textrm{diff}}(r)\propto r^{-4.03 \pm
0.15}$, as predicted by Eq.~(\ref{Ir}) for $\alpha=1.01\pm0.04$.  Moreover, this asymptotic behavior is valid for a wide
range of opacities as shown on Fig.~\ref{fig5}\,(c). We note that for increasing opacity, the total relevant signal is
reduced [see Fig.~\ref{fig4}] and the relative importance of amplified spontaneous emission pedestal increases,
requiring thus to adapt the range of radial distances where a reasonable fit can be obtained.  Finally, we stress that
the scaling laws in the tail of $T_{\textrm{diff}}(r)$ can be obtained from a single image on the CCD, giving access to
L\'evy parameter with a single snapshot whose duration is $\unit{2}{\milli\second}$. This has to be compared to previous
studies where the L\'evy parameter were extracted in $\unit{30}{\hour}$~\cite{MERCADIER-2013}. Also, with only one
snapshot, the vertical dynamics is more than four orders of magnitude and allow a good fit. Furthermore, this single
image under multiple scattering condition allows to make use of these scaling laws in different experimental conditions,
in laboratory environment or in astrophysical systems.

One interesting point concerning the radial profile is the behavior at small distances $r$. In L\'evy
glasses~\cite{WIERSMA-2008-1}, a pronounced cusp is observed close to $r=0$ which is not the case neither in our
experiment nor in simulations. There are several fundamental differences between L\'evy glasses and hot atomic vapors
illuminated by a monochromatic laser beam. For L\'evy glasses, when light enters the system, the step-size distribution
is already a power law with an exponent $\alpha<2$ leading to superdiffusion. In our hot vapor experiment in contrast,
the correct step-size distribution for photons at the laser frequency is an exponential function
$\mu_{\ell}\left(\delta_0\right)\exp\left[-\mu_{\ell}\left(\delta_0\right)x\right]$, different from the power law
step-size distribution after one scattering event, where the frequency is already redistributed. We have checked
numerically that if a broadband excitation were used to illuminate the hot atomic vapor cell, a sharp cusp around the
center of the transmitted light path is recovered.

\section{Conclusion}

In summary, we have shown theoretical, numerical and experimental results confirming L\'evy flights of photons in hot
vapor under multiple scattering condition with annealed disorder. L\'evy flights in our system appear due to the
interplay between the high quality factor of atomic resonances and large Doppler broadening. The scaling laws obtained
for the total diffuse transmission $T_{\textrm{diff}}$ and intensity profile $T_{\textrm{diff}}(r)$ suggests that
other spectral broadenings such as collisions with a buffer gas could allow to enhance or control L\'evy flights and
will be accessible in realistic experimental conditions. This work opens also new perspectives to look for evidence of
L\'evy flight in astrophysical systems or for quantitative studies of the limits of validity of the complete frequency
redistribution in hot vapors.

\begin{acknowledgements}
   We thank Dominique Delande for fruitful discussions and we acknowledge funding for N.M. and Q.B. by the french
   Direction G\'en\'erale de l'Armement. R.P acknowledges the support of LABEX WIFI (Laboratory of Excellence
   ANR-10-LABX-24) within the French Program ``Investments for the Future'' under reference ANR-10-IDEX-0001-02
   PSL$^{\ast}$. E.J.N. acknowledges FCT (441.00 CNRS).
\end{acknowledgements}

\section{Appendices}

\subsection{Monte Carlo scheme}\label{monte_carlo}

The exact numerical resolution of Eq.~(\ref{etr_levy}) is possible using a Monte Carlo scheme. As a Monte Carlo
algorithm is designed to evaluate numerically complex integrals, a new form of the integro-differential transport
equation has to be written. First, we take the Fourier transform of Eq.~(\ref{etr_levy}) with respect to the space and
time variables which gives
\begin{multline}\label{etr_levy_tf}
   \left[i\bm{u}\cdot\bm{q}+\left(1-\frac{i\Omega}{\Gamma}\right)\mu_{\ell}\left(\delta\right)\right]I\left(\bm{u},\bm{q},\delta,\Omega\right)
\\
      =\frac{\mu_{\ell}\left(\delta\right)}{4\pi}\int p\left(\delta,\delta',\bm{u},\bm{u}'\right)I\left(\bm{u}',\bm{q},\delta',\Omega\right)
         \ud\bm{u}'\ud\delta'.
\end{multline}
Using this expression, we can easily isolate the Fourier transform of the specific intensity
$I\left(\bm{u},\bm{q},\delta,\Omega\right)$. Noticing that
\begin{multline}
   \left[i\bm{u}\cdot\bm{q}+\left(1-\frac{i\Omega}{\Gamma}\right)\mu_{\ell}\left(\delta\right)\right]^{-1}
\\
      =\int_0^{\infty}\exp\left[-\left\{i\bm{u}\cdot\bm{q}+\left(1-\frac{i\Omega}{\Gamma}\right)\mu_{\ell}\left(\delta\right)\right\}s\right]\ud s
\end{multline}
we obtain
\begin{multline}
   I\left(\bm{u},\bm{q},\delta,\Omega\right)=\frac{\mu_{\ell}\left(\delta\right)}{4\pi}
\\
      \times\int_{s=0}^{\infty}\exp\left[-\left\{i\bm{u}\cdot\bm{q}+\left(1-\frac{i\Omega}{\Gamma}\right)\mu_{\ell}\left(\delta\right)\right\}s\right]
\\
      \times\int p\left(\delta,\delta',\bm{u},\bm{u}'\right)I\left(\bm{u}',\bm{q},\delta',\Omega\right)\ud\bm{u}'\ud\delta'\ud s.
\end{multline}
Finally, by taking the inverse Fourier transform of the last expression, we find
\begin{multline}
   I\left(\bm{u},\bm{r},\delta,t\right)=\int_{s=0}^{\infty}\mu_{\ell}\left(\delta\right)\exp\left[-\mu_{\ell}\left(\delta\right)s\right]
\\
      \times\int \frac{p\left(\delta,\delta',\bm{u},\bm{u}'\right)}{4\pi}
         I\left(\bm{u}',\bm{r}-s\bm{u},\delta',t-\frac{\mu_{\ell}\left(\delta\right)s}{\Gamma}\right)\ud\bm{u}'\ud\delta'\ud s.
\end{multline}
This form is very convenient to derive the Monte Carlo algorithm dedicated to the resolution of Eq.~(\ref{etr_levy}). In
particular, it shows that the probability density to perform a step of length $s$ between two scattering events is given
by $\mu_{\ell}\left(\delta\right)\exp\left[-\mu_{\ell}\left(\delta\right)s\right]$ and the probability densities to be
scattered in a direction $\bm{u}$ from a direction $\bm{u}'$ and at a frequency $\delta$ from a frequency $\delta'$ are
respectively given by $1/\left(4\pi\right)$ and $p\left(\delta,\delta',\bm{u},\bm{u}'\right)$. As the probability
densities are known analytically, no truncation is performed and the method of resolution is exact, the numerical error
being given by the number of energy packets sent into the system.

\subsection{Standard diffusion radial profiles}\label{appendix-diffusion}

This appendix is dedicated to the technicalities of the derivation of the asymptotic behavior of the radial profile of
the transmitted or reflected intensity in the case of classical diffusion. The diffusion equation is solved for an
infinite slab, in steady state, using two canonical approximations: the source term is substituted for a unit delta
production (at a transport mean-free path $\ell^*$ from left side; excitation is assumed to impinge from left) and
Dirichlet boundary conditions, defined at a so called extrapolation length ($z_{e}$; the effective width is
$L_{e}=L+2z_{e}$, with $L$ the slab width)~\cite{ZHU-1991}. The problem to solve is
$-D\nabla^{2}I\left(\bm{r}\right)=\delta\left(\bm{r}\right)$, with $I\left(z=0\right)=I\left(z=L_{e}\right)=0$ (the
origin of the~$z$ coordinate makes the application of the theory of Green functions --fundamental Green function and
multiple image series or, alternatively, the expansion in eigenfunctions-- the simplest possible). Two further
quantities are defined: the flux $\Phi\equiv-D\partial I\left(\bm{r}\right)/\partial z$ and the radial profiles, in
reflection and transmission,
$\left\{R_{\textrm{diff}}\left(r\right)=-\Phi\left(z=z_{e}\right),T_{\textrm{diff}}\left(r\right)=\Phi\left(z=L+z_{e}\right)\right\}$,
here assumed with axial symmetry (experimental case). It is assumed no absorption. We only use the transmitted profile,
but quote also the one in reflection, for reference.

The solution in Fourier space is known~\cite{BARTHELEMY-2009} and is given by
\begin{align}
      R_{\textrm{diff}}\left(q_{\bot}\right) & = \frac{1}{2\pi}
         \frac{\cosh\left(q_{\bot}z_{e}\right)
         \sinh\left(q_{\bot}\left(L_{e}-\left(\ell+z_{e}\right)\right)\right)}{\sinh\left(q_{\bot}L_{e}\right)},
   \\
      T_{\textrm{diff}}\left(q_{\bot}\right) & = \frac{1}{2\pi}
         \frac{\cosh\left(q_{\bot}z_{e}\right)i
         \sinh\left(q_{\bot}\left(\ell+z_{e}\right)\right)}{\sinh\left(q_{\bot}L_{e}\right)},
\end{align}
where $q_{\bot}$ is the Fourier conjugate of the radial distance (the $2\pi$ is dependent on Fourier notation choice).
The solution in physical space is obtained by performing the inverse Fourier transform, as usual. Care must be exercised
in two points: it's a 2D Fourier inverse transform and the implementation of the residues theorem must take into account
the asymptotic nontrivial behavior of the Bessel function. The final solution in physical space is given
by Eqs.~(\ref{diff_eq_solution R}) and~(\ref{diff_eq_solution}).

Its derivation used the facts~\cite{BADDOUR-2011,ABRAMOWITZ-1972}:
\begin{enumerate}
   \item the 2D Fourier transforms of axial symmetric functions are the Hankel transforms of zero order;
   \item the Bessel function can be written as
   \begin{equation}
      \operatorname{J}_{n}\left(z\right)=\frac{1}{2}\left[\operatorname{H}_{n}^{\left(1\right)}\left(z\right)
         +\operatorname{H}_{n}^{\left(2\right)}\left(z\right)\right];
   \end{equation}
   \item the special Hankel functions asymptotics are
   \\$\left\{\operatorname{H}_{n}^{\left(1\right)}\left(z\right)\rightarrow 0,\,\left|z\right|\rightarrow\infty\,\&\,0<\arg\left(z\right)<\pi,\right.$
   $\left.\operatorname{H}_{n}^{\left(2\right)}\left(z\right)\rightarrow0,\,\left|z\right|\rightarrow\infty\,\&\,-\pi<\arg\left(z\right)<0\right\}$.
\end{enumerate}

\subsection{Fitting details}\label{appendix-fitting-details}

This appendix is dedicated to several aspects of the various fitting procedures, used in the main body of the paper.

The correct use of the analytical representations of the radial profiles in Eqs.~(\ref{diff_eq_solution R})
and~(\ref{diff_eq_solution}) requires special care near the origin. The Hankel functions in these equations diverge at the
origin. We do not have an analytical proof of the series convergence. However, we made a detailed study of the
convergence speed, near the origin and found empirical evidence of the convergence:~in all cases, using a sufficiently
high number of terms, we verified the convergence of the series. We nevertheless emphasize that the convergence is very
slow:~by values of the argument of the order of~$\power{10}{-5}$, we needed about~$40000$ terms to warrant convergence.
All the fits in this work used a fixed number of terms in the series ($2500$, a conservative estimate, based in the
smallest distance we used).

The effective use of Eqs.~(\ref{diff_eq_solution R}) and~(\ref{diff_eq_solution}) in fitting routines requires care in at
least two details. The first was already mentioned:~check convergence, at the smallest distance. The second is connected
with the limitations embedded by the physics. If $\ell^*\ll L$, then $L_{e}\sim L$ and the exponential asymptotic does not
depend upon the transport mean-free path. In fact, the normalized transmitted profile is effectively constant in this well
developed diffusion regime, thus reducing the ability to recover meaningfully parameters by
fitting Eqs.~(\ref{diff_eq_solution}) to experimental data. This of course is what is expected from simple physical
arguments. In these cases, a better approach is to devise an experimental setup in which both reflection and
transmission profiles are accessible at the same time. A global analysis will have~$\ell^*$ sensible fitted from the
central part of the reflection, and~$L$ from the asymptotic of both reflection and transmission.

A detailed parameter space exploration of Eqs.~(\ref{diff_eq_solution R}) and~(\ref{diff_eq_solution}), to judge whether
one can expect artifacts, is not the focus of this work. However, we checked \emph{a posteriori} and indeed, the
transmission profile is quite insensitive to the actual value of the $\ell^*$ value, in the fits of Fig.~\ref{fig2}.
This can also be quantified in a computational linear algebra approach by the actual value of the condition number of
the fitting matrix, the ratio of biggest to smallest eigenvalue of a decomposition of the Hessian matrix of the fit. In
the fits of Fig.~\ref{fig2}, the condition number is of the order of $\power{10}{-9}$. This is a very small value thus
signaling instrinsic numerical limitations. Given all these arguments, the agreement between Monte Carlo input
parameters $\ell^*=\unit{0.05}{\milli\meter}$, $L=\unit{14.82}{\milli\meter}$) and the ones recovered by
fitting ($\ell^*=\unit{0.09 \pm 2.5}{\milli\meter}$, $L=\unit{14.7 \pm 3.5}{\milli\meter}$) is very good. 

An important detail connected with the procedures used in this work for statistical inference should be highlighted. It is well
known that fitting to a power law is notoriously difficult. This of course results from the fact that, in order to have 
meaningful fitted values of the power law exponent, one must include in the statistical inference values spanning several
orders of magnitude (in principle, as many as possible). Two general procedures are then available: either data analyse binned
of unbinned information. The unbinned information is preferable in theoretical grounds since the binning of data can make
the estimators biased. The use of maximum likelihood estimators for unbinned data is far less prone to 
artefacts~\cite{GOLDSTEIN-2004}. This is the cause of the use of unbinned data in such classical works 
as~\cite{VISWANATHAN-1996, EDWARDS-2007}, where animal foraging 
movements are tracked. However, one of the strongest arguments of the experimental study we present in this work is the very
high signal to noise ratio we are able to achieve by analysing CCD images. It is as if each individual photon corresponded 
to a single event, in animal tracking studies. We have in a single~$\unit{2}{\milli\second}$ snapshot probably more information than all the data 
collected thus far in all the animal tracking studies realized thus far by humankind. By using a CCD, the data is already 
recorded in a binned (in each pixel) format. And we obtain very high signal to noise ratios, spanning as much as four orders of 
magnitude in dynamical range. We do not have access to the unbinned data and can not check for artefacts the data analysis 
we used (we used Levenberg-Marquardt nonlinear least squares algotrithms, which are also unbiased maximum likelihood estimators,
for a Gaussian distribution of the signal in each pixel of the camera), by comparing the results obtained for unbinned data
analysis. We nevertheless made your best efforts to dismiss 
fitting artefacts: we fitted the data in Fig.~\ref{fig2} using nonlinear least squares, unweighted in linear scale as well as 
weighted (Poisson counting, weight due to change of independent variable into a logarithmic scale), using linear least squares in a log-log scale and also testing the sensitivity of the actual fitted parameters to the initial approximations. In all cases, 
the differences in the power law parameter was smaller than the confidence intervals we quote in this work.

\end{document}